\documentclass[pra,twocolumn,superscriptaddress,nobibnotes,letterpaper]{revtex4-1}

\usepackage[pdftex]{graphicx}
\usepackage[pdftex]{epsfig}
\usepackage{amsmath}
\usepackage{amssymb}
\usepackage{amsfonts}
\usepackage{color}
\usepackage{wrapfig}
\usepackage{eucal}
\usepackage{hhline}
\usepackage{soul}

\usepackage[table]{xcolor}
\usepackage{array}
\definecolor{Gray}{gray}{0.8}

\begin{document}

\pagenumbering{arabic}

\title{Platform for measurements of the Casimir force between two superconductors}\thanks{This work was published in Phys.\ Rev.\ Lett.\ \textbf{121}, 030405 (2018).}

\author{R.\ A.\ Norte} \email{r.a.norte@tudelft.nl}
\affiliation{Department of Quantum Nanoscience, Kavli Institute of Nanoscience, Delft University of Technology, Lorentzweg 1, 2628CJ Delft, The Netherlands}
\author{M. Forsch}
\affiliation{Department of Quantum Nanoscience, Kavli Institute of Nanoscience, Delft University of Technology, Lorentzweg 1, 2628CJ Delft, The Netherlands}
\author{A. Wallucks}
\affiliation{Department of Quantum Nanoscience, Kavli Institute of Nanoscience, Delft University of Technology, Lorentzweg 1, 2628CJ Delft, The Netherlands}
\author{I. Marinkovi\'{c}}
\affiliation{Department of Quantum Nanoscience, Kavli Institute of Nanoscience, Delft University of Technology, Lorentzweg 1, 2628CJ Delft, The Netherlands}
\author{S.\ Gr\"oblacher} \email{s.groeblacher@tudelft.nl}
\affiliation{Department of Quantum Nanoscience, Kavli Institute of Nanoscience, Delft University of Technology, Lorentzweg 1, 2628CJ Delft, The Netherlands}


\begin{abstract}
Several experimental demonstrations of the Casimir force between two closely spaced bodies have been realized over the past two decades. Extending the theory to incorporate the behavior of the force between two superconducting films close to their transition temperature has resulted in competing predictions. To date, no experiment exists that can test these theories, partly due to the difficulty in aligning two superconductors in close proximity, while still allowing for a temperature-independent readout of the arising force between them. Here we present an on-chip platform based on an optomechanical cavity in combination with a grounded superconducting capacitor, which overcomes these challenges and opens up the possibility to probe modifications to the Casimir effect between two closely spaced, freestanding superconductors as they transition into a superconducting state. We also perform preliminary force measurements that demonstrate the capability of these devices to probe the interplay between two widely measured quantum effects:\ Casimir forces and superconductivity.
\end{abstract}

\maketitle

The Casimir force can be interpreted as a direct consequence of the quantization of the electromagnetic field. It was first postulated by Casimir~\cite{Casimir1948a,Casimir1948b} in 1948 and conclusively experimentally observed by Lamoreaux in 1997~\cite{Lamoreaux1997}. As two conducting plates are placed at a small distance from one another, an attractive force manifests between them. The effect was originally suggested to result from quantum vacuum fluctuations, while later derivations by Lifshitz explain its emergence from charge fluctuations in the conductors~\cite{Lamoreaux2007}. While the Casimir effect has been tested with various metallic and dielectric systems, it remains an open question how the force behaves between closely spaced surfaces as they transition into a superconducting state. It is known that superconductivity affects the reflectivity of a metal at frequencies below $k_{\mathrm{B}}T_\mathrm{c}/\hbar$ and that it should also change how the magnitude of the Casimir force depends on the temperature above and below the superconducting transition temperature $T_\mathrm{c}$~\cite{TinkhamTc1957,Bimonte2005a,Klimchitskaya2017,EerkensPhD}. By experimentally determining these corrections around $T_\mathrm{c}$, it should be possible to distinguish between competing descriptions of the frequency response of superconductors, such as the Drude and plasma models, as appropriate descriptions for the TE zero mode in calculations of the Casimir effect~\cite{Bimonte2008}. In addition, measuring the force arising between two superconductors could allow us to test recent proposals of a potential gravitational Casimir effect~\cite{Quach2015,Calloni2016}.

One of the main obstacles for such Casimir experiments has been the difficulty to realize small gaps between the surfaces of two objects at low temperatures~\cite{laurent2012casimir,castillo2013casimir}. Sophisticated nanopositioning systems are typically required to place the two objects tens of nanometers apart. While scanning tunneling microscopes can commonly achieve subnanometer spacings between a tip and a surface at a single atomic point, this type of proximity becomes difficult over the large surfaces and small gap sizes required to probe a measurable Casimir force. Many experiments have therefore used a sphere close to a plate for easier alignment (sphere-plate geometry), which significantly reduces the force compared to the original proposal of using two parallel plates. In addition, robust detection systems are required to not only measure the forces between these objects, but also to modulate and stabilize the gap to nanometer precision~\cite{Harris2000b,Decca2005,Chan2008,Garcia-Sanchez2012,Zou2013}.

Incorporating superconductors introduces a number of additional challenges. For example, realizing a stable force detection that operates over a large temperature range, ideally well above and below the $T_\mathrm{c}$ of a superconductor, is difficult. A recent experiment using an electromechanical system was able to infer the Casimir force exclusively below $T_\mathrm{c}$, where a superconducting LC readout circuit operates without significant losses~\cite{Andrews2015}. It is also crucial that the force detection does not compromise the superconductivity of the films through additional heating by optical absorption, magnetic fields or currents from measurements. Verifying that the entire structure is superconducting requires an integrated way to monitor the plates' resistance.

\begin{figure}[t!]
\begin{center}
\includegraphics[width=\columnwidth]{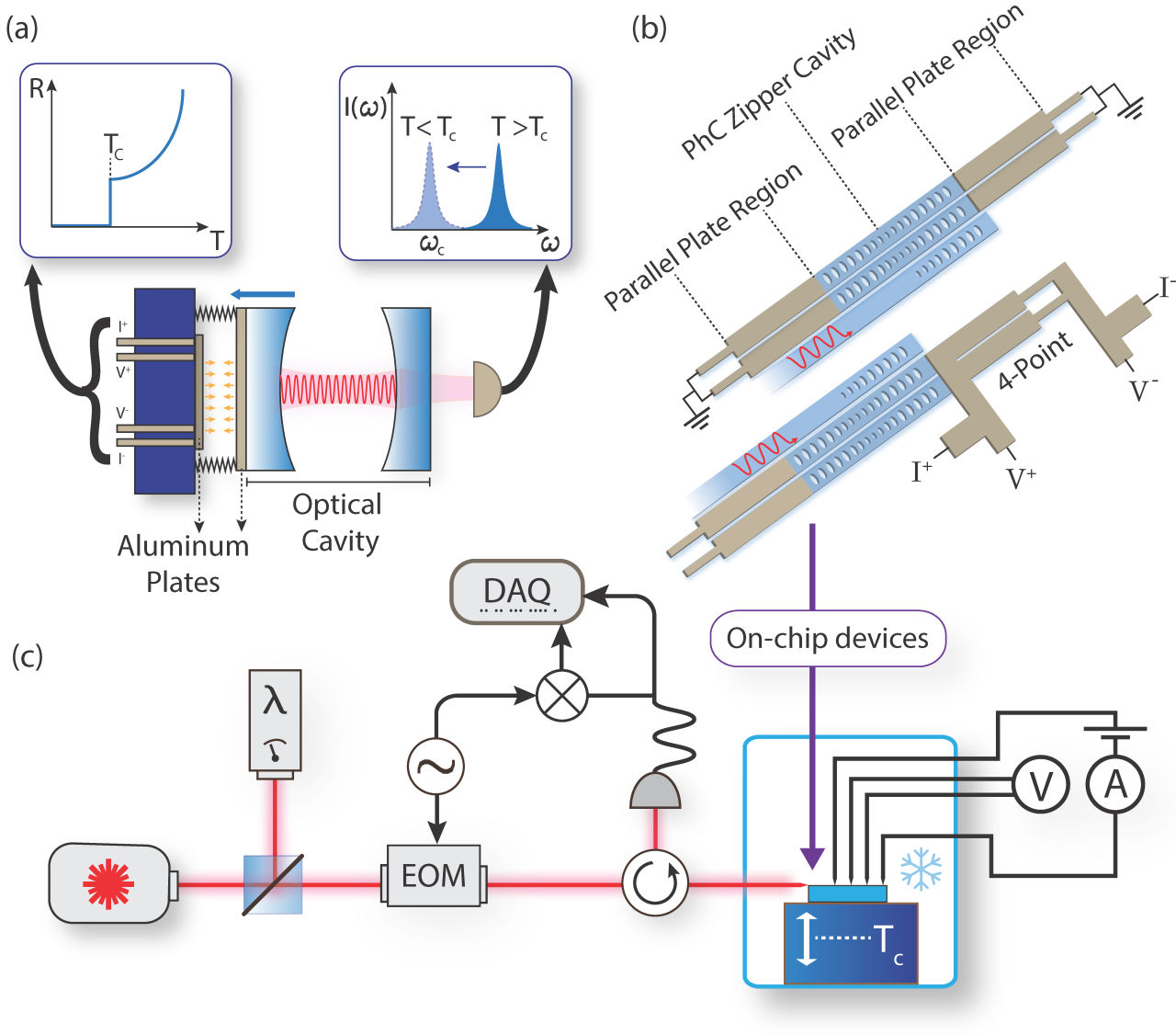}
\caption{(a) Working principle of the platform:\ The movable mirror of an optomechanical cavity is coated with an aluminum film and positioned in close proximity to a fixed aluminum plate. Any forces that arise between the plates at the onset of superconductivity would change the position of the movable mirror, which can be measured through a shift in the optical cavity resonance. (b) Illustration of our on-chip implementation of (a). We design a zipper cavity made of silicon nitride and evaporate aluminum on its support strings. By evanescently coupling light through a waveguide to the cavity we can monitor its resonance frequency as a function of temperature. The resistance of the film is continuously probed through an additional device on the chip that is connected to a four-point measurement (bottom), allowing us to determine its $T_\mathrm{c}$. (c) Sketch of the setup. We stabilize our laser to the resonance of the optomechanical cavity using a wavelength meter. The resonance frequency is monitored through a phase measurement by modulating sidebands onto the light field using an electro-optical modulator (EOM), which are detected, mixed down and then digitized. The sample itself is mounted in a dilution refrigerator, where we can ramp the temperature above and below $T_\mathrm{c}$, while monitoring its resistivity with the four-point measurement.} 
\label{fig:1}
\end{center} 
\end{figure} 

Here we realize a novel on-chip optomechanical sensor that allows us to optically measure the changes in the force between two plates made of superconducting material at any temperature, in particular while they are cooled through $T_\mathrm{c}$. By engineering the entire superconducting structure and nanophotonic detection system on chip, we realize a versatile measurement platform that can be readily used inside a dilution refrigerator at Millikelvin temperatures. The integrated photonic crystal microcavities allow us to measure modifications to the Casimir force before and after the onset of superconductivity with a resolution of 6~mPa between the surfaces. We have developed new fabrication methods that utilize high-stress films to realize state-of-the-art parallelism between freestanding superconducting surfaces over long distances without active stabilization. Our platform also has on-chip circuitry, which we use to accurately determine the superconductors' properties such as resistance, penetration depth, and coherence length. The optical force detection is stable at low temperatures, easily coupled using fiber optics, and operates at all temperatures without compromising the superconductivity of the films.

A generic schematic of our experiment is shown in Fig.~\ref{fig:1}(a). It consists of two parallel plates of a superconductor, which we choose to be aluminum. One of these plates is attached to the movable mirror of an optomechanical cavity. We monitor the resistance of the superconducting plates using a four-point measurement as they are cooled below their superconducting transition temperature $T_\mathrm{c}$. Any temperature-dependent forces will affect the distance between the plates, causing an effective change in length of the cavity and a shift in the cavity resonance frequency $\omega_\mathrm{c}$.

\begin{figure}[h!]
	\begin{center}
		\includegraphics[width=\columnwidth]{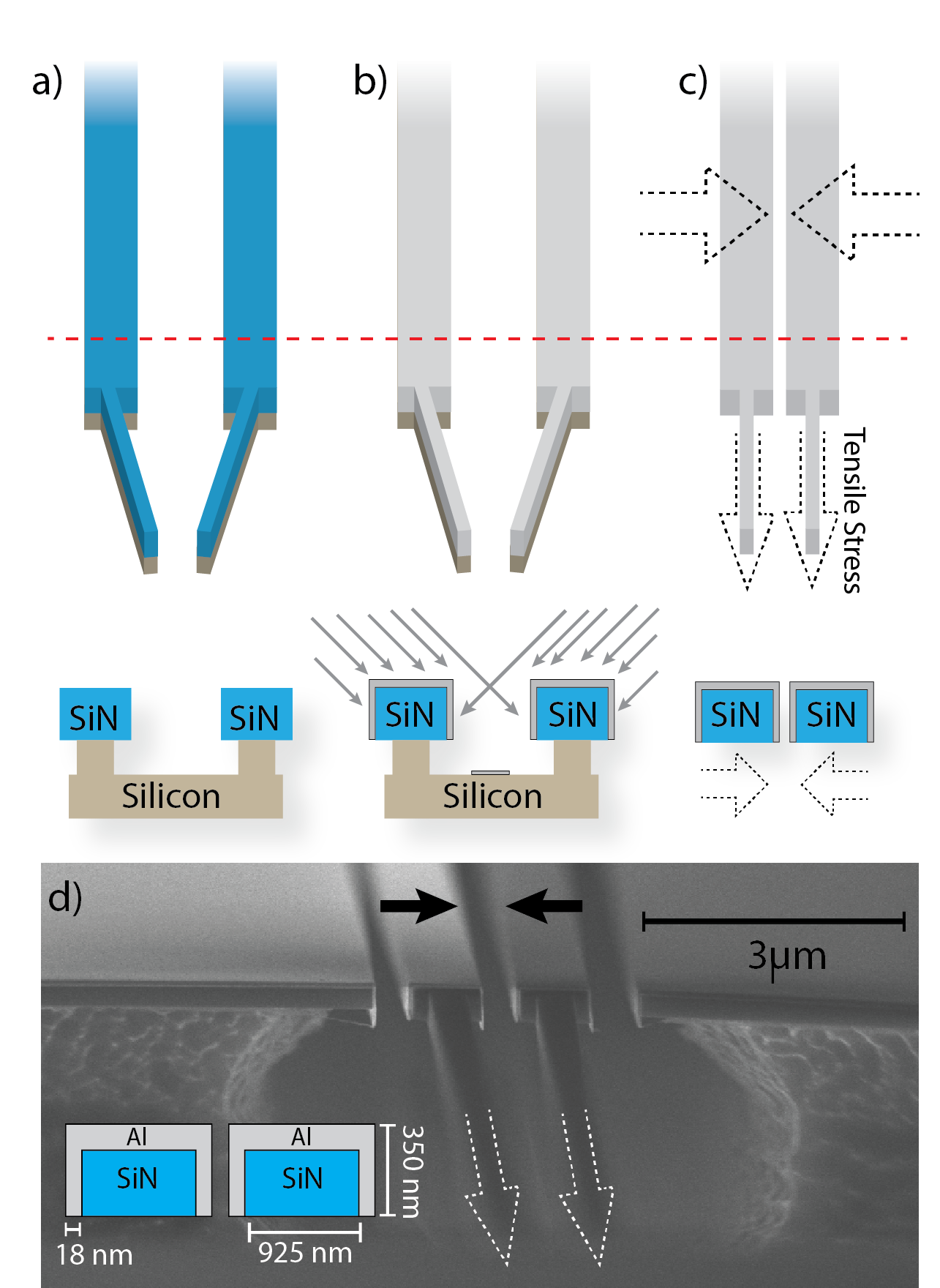}
		\caption{We highlight how tensile stress in the SiN films is utilized to achieve small gaps between parallel superconducting plates over long distances. The bottom insets show the corresponding cross section through the top nanobeams (indicated by the red dashed line). (a) First, nanobeams are etched into the SiN film and into the Si substrate underneath. (b) Aluminum is then evaporated onto the nanobeams at two different angles to homogeneously cover the beams on its sides in order to form plates. (c) After the Si underneath the SiN is removed, the beams are free to move. The tensile stress in the SiN pulls the strings straight and taut. This simultaneously moves the nanobeams closer together, forming small gaps. The initial lithographically defined angle determines the gap size. (d) Side-view SEM image showing the profile of the metal that makes up the plates for the measurements. Here we show a device designed to have a 300~nm gap once suspended. The inset shows a cross section of the metallized portions of the nanobeams.}
		\label{fig:2}
	\end{center} 
\end{figure}

\begin{figure}[t!]
	\begin{center}
		\includegraphics[width=\columnwidth]{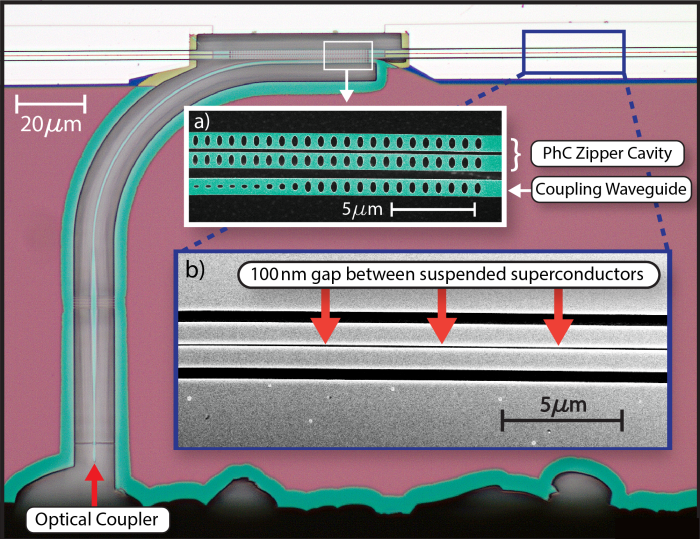}
		\caption{Microscope image of a zipper cavity with integrated superconducting strings. In the lower left corner is an adiabatic coupler used to couple light from a lensed fiber to a waveguide, which is then evanescently coupled to the suspended photonic crystal zipper cavity. A close-up of this nanophotonic structure is shown in inset (a), where the reflector at the end of the waveguide is clearly visible, which allows us to send the reflected laser light back into the fiber. Inset (b) is a scanning electron microscope image of a section of the $170~\mu m$ suspended nanostrings, which are partially covered in aluminum and which, due to the large tensile stress in the silicon nitride film, are almost perfectly parallel with a gap of only 100~nm.}
		\label{fig:3}
	\end{center} 
\end{figure}

We realize an optomechanical system equivalent to the sketch in Figure~\ref{fig:1}(a) by etching a photonic crystal cavity into two adjacent, high-stress (1.3~GPa) LPCVD silicon nitride (SiN) beams, usually referred to as a zipper cavity~\cite{Krause2015} [see Figs.~\ref{fig:1}(b), \ref{fig:2}, and~\ref{fig:3} for more details]. Any relative motion between the two beams results in a shift of the optical cavity frequency $\omega_\mathrm{c}$. Suspending the photonic crystals are long tethers which connect to the substrate. The overall structure consists of two 384~$\mu$m-long parallel strings with a width $w = 926$~nm and thickness $t = 300$~nm. The large tensile stress allows the parallel beams to remain straight and closely spaced over long distances~\cite{Norte2014}. The tethers are partly covered with an aluminum film, which forms a pair of parallel plates. They are deposited with an 18~nm effective thickness (the thickness of the aluminum deposited on the sides of the nanostrings) and surface area $A = 220~\mu$m $\times$ 350~nm [cf.\ inset in Fig.~\ref{fig:2}(d)]. Unlike conventional Casimir measurements, we realize a force measurement with lithographically defined gaps in order to avoid spurious heating of the superconducting film from actively stabilizing the gap size. This allows us to have several devices per chip, each with a different predetermined gap size, with an achievable gap as small as $\sim$100~nm.

In order to fabricate the devices, we first etch the nanobeam structures into the SiN films and then proceed to etch deeper into the silicon substrate to produce the profile seen in Fig.~\ref{fig:2}(a). The initial large spacing between the nanobeams is important for two reasons:\ First, it allows us to make straight and vertical sidewalls of the SiN using a directional plasma etch (CHF$_3$/O$_2$). Lithographically defining and etching small gaps can be prone to surface roughness and angled sidewalls, which can significantly limit the achievable gaps between the metallized beams. Second, the large gaps allow for enough space between the beams to homogeneously cover the sidewalls with aluminum using an angled-stage electron-beam evaporator [cf.\ Fig.~\ref{fig:2}(b)]. The Si below the beams is then removed using an SF$_6$ plasma release. Once free to move, the tensile stress in the beams pulls them straight and taut, simultaneously also pulling them closer together [Fig.~\ref{fig:2}(c)]. The final gap size between the beams is defined by the initial lithographic angle of the SiN tethers seen in Fig.~\ref{fig:2}(a). By sweeping the angle of these tethers over several devices, we can controllably realize a variety of gap sizes on a chip. These pull-in techniques allow us to engineer the high-tensile stress in SiN films to achieve excellent parallelism over long distances~\cite{Norte2014}.

To avoid spurious shifts from capacitive forces and minimize the influence of patch potentials~\cite{Speake2003}, all wires are grounded. At ambient temperatures the initial gap between the nanostrings already includes the Casimir force. The zipper cavity is located in the center of the beam where the expected displacement is maximal. The optical measurement of the cavity's resonance frequency is done by coupling laser light through a lensed fiber to a waveguide that is brought into close proximity of the cavity and hence allows us to evanescently couple light in and out with a total efficiency of about 55\% [see Fig.~\ref{fig:1}(b), the SI, and Ref.~\cite{Groeblacher2013a,Riedinger2016} for details]. The light is then reflected back into the fiber, where we send it through an optical circulator and into a photodetector.

\begin{figure*}[t!]
	\begin{center}
		\includegraphics[width=2\columnwidth]{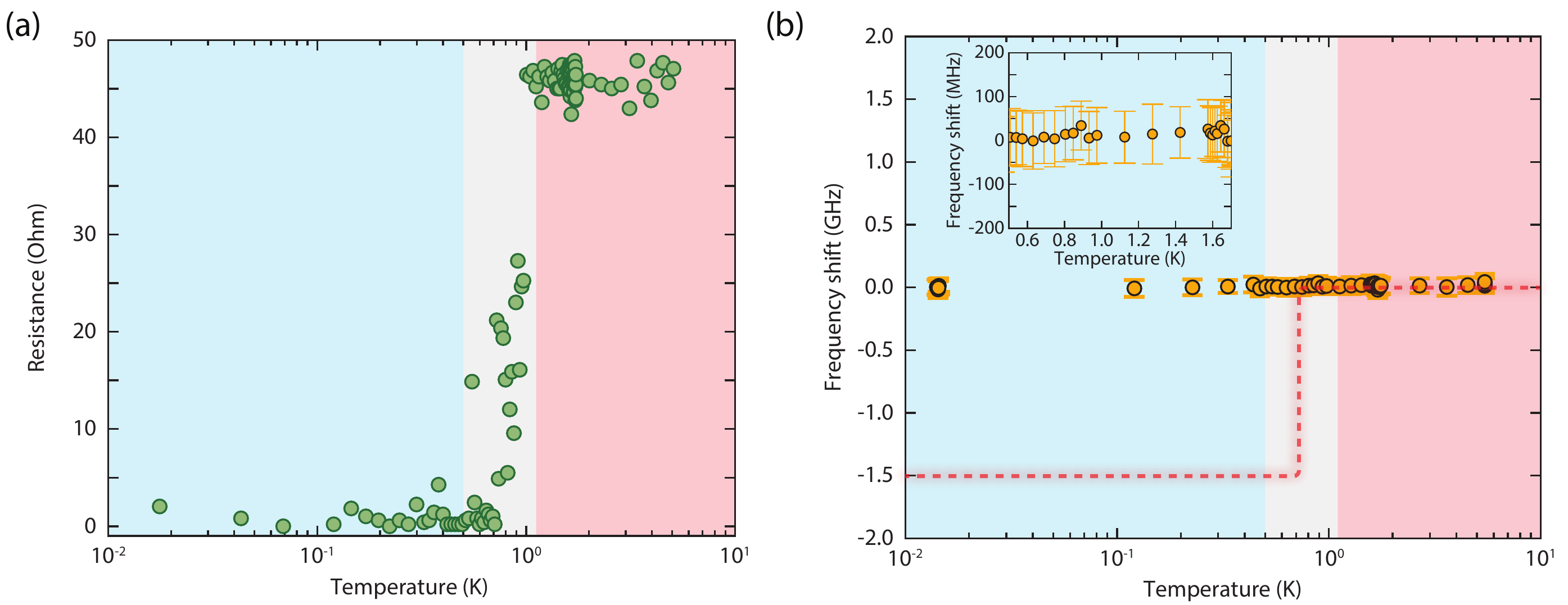}
		\caption{(a) Shown is the resistance of the superconducting wires on our test device as a function of temperature. A jump from a few $\Omega$ to about 45~$\Omega$ is clearly visible just below 1~K. The data are collected from several four-point measurements using 1~$\mu$A of current, which results in a stable and very reproducible curve. The shaded regions are a guide to the eye, with blue representing the superconducting, gray the transition, and red the normal conducting state of the Al wire. (b) We plot the relative change in resonance frequency of the fundamental mode of a representative zipper cavity as a function of temperature. A force acting on the wires when they are in their superconducting state would shift the resonance frequency in the transition region between normal conductance and superconductance [see Fig.~\ref{fig:1}(a) for more details]. The gap size for this particular device is 100~nm ($\pm10$~nm, see SI) with a measured optomechanical coupling rate of 50~GHz/nm. The red dotted line depicts the expected shift in frequency for a recently proposed gravitational Casimir experiment (see the text). The inset shows an overall variation in the central frequency of less than 30~MHz, mostly coming from variations and drifts in our measurement setup. Each data point is the average of 100 samples, and the error bars are shown in s.d.}
		\label{fig:4}
	\end{center}
\end{figure*} 

We also fabricate a reference device with the same dimensions, which allows us to perform four-point measurements of the aluminum wire resistance and hence accurately determine when the wires become superconducting using a low-current ($\sim$$\mu$A) source [see Fig.~\ref{fig:1}(b)]. The measurements are performed in a closed-cycle dilution refrigerator with a base temperature of 10~mK, allowing us to cool the aluminum well below its $T_\mathrm{c}$ of approximately 1.3~K. The actual critical temperature for our devices is measured to be slightly below 1~K [cf.\ Fig.~\ref{fig:4}(a)] -- we attribute this deviation from the literature value of aluminum to our particular geometry, which features thin wires~\cite{Romijn1982} and minimal thermal conductivity due to its freely suspended SiN support.

When we first cool the zipper cavities from room temperature to about 3~K, we observe a shift in the optical resonance of typically a few nanometers, which is due to the temperature-dependent refractive index and thermal contractions. Most of the shift happens above 10~K, and no further shift in resonance frequency is expected or observed below that temperature. This is due to both silicon nitride and aluminum having a coefficient of linear thermal expansion approaching zero at cryogenic temperatures~\cite{Pobell2007}. At first, we measure the device that has its aluminum wires connected to the four-point probe. At base temperature (10~mK), we continuously monitor the resistance as a function of the optical power sent into the zipper cavity. We use this measurement to choose a laser power for the actual measurements, 200~nW at the input of the dilution refrigerator, that is well below the threshold for the superconductivity to break down due to absorption of the laser and subsequent heating of the metal ($\sim1~\mu$W).

We then perform measurements on several devices with gap sizes ranging from $a = 300$~nm down to just below $a = 100$~nm [for a sketch of the setup, see Fig.~\ref{fig:1}(c)]. The results for the bigger gaps do not deviate from the ones for the smaller gaps, and we will only focus on $a = 100$~nm, which should also exhibit the largest effects. The device used to obtain the data in Fig.~\ref{fig:4}(b) has a fundamental cavity resonance at $\lambda = 1586.3~$nm ($\omega_\mathrm{c}\approx 2\pi\times 189$~THz) and a linewidth of $\kappa \approx \omega_\mathrm{c}/Q_o = 2\pi\times 4.2$~GHz (optical quality factor of $Q_\mathrm{o}=4.5\times 10^4$). We experimentally determine our coupling strength, which is the frequency shift of the cavity as a function of displacement, to be $g_{\mathrm{OM}}/2 \pi \approx 50~$GHz/nm (see SI), in good agreement with our simulated value of $50~$GHz/nm. In our experiment we detect frequency shifts of the cavity resonance resulting form the Casimir force by measuring a phase shift using a Pound-Drever-Hall scheme [c.f.\ Fig.~\ref{fig:1}(b) and the SI for details], with a minimally resolvable shift of 10~MHz (around 0.25\% of the optical linewidth).

The state-of-the-art parallelism we have developed here gives us a minimal resolvable Casimir pressure of 6~mPa, corresponding to a gap change of the zipper cavity of a few hundred femtometers. When changing the temperature through the $T_\mathrm{c}$ of aluminum, we do not detect any frequency shift, as can be seen in Figure~\ref{fig:4}(b). We repeat this measurement various times, with optical powers well below and above the critical power for the superconductivity to break down, and do not observe a shift. Our sensitivity already allows us to disprove the validity of a gravitational Casimir force predicted to occur between superconductors~\cite{Quach2015}. For our device, such forces are estimated to be 0.5~Pa, resulting in a cavity frequency shift of 1.5~GHz. One potential explanation for the absence of this Casimir force in our data is that the particular choice of our geometry could have a reducing effect on the force between the two superconducting wires. All our calculations are done for plates, which could in principle deviate from the actual geometry that we fabricate due to edge effects. In order to estimate the reductions in the Casimir pressure due to our parallel nanobeam geometry (compared to the infinite parallel plates) we turned to numerical calculations performed in ~\cite{Pernice2010} which use finite-difference time-domain methods~\cite{rodriguez2009casimir,mccauley2010casimir} to calculate the Casimir force between two suspended rectangular beams at zero temperature -- in good agreement with proximity force approximations~\cite{derjaguin1956direct} of the same geometries. For experiments testing the Casimir force with similar devices, simulations and experiments have found the force to be about an order of magnitude smaller than the ideal plate-plate calculations~\cite{Pernice2010}. However, the gravitational Casimir force would have to be about 2 orders of magnitude smaller than the large plates calculation for it to be undetectable in our current measurement [see inset in Fig.~\ref{fig:4}(b)].

To conclude, we have designed and fabricated on-chip optomechanical sensors that allow us to probe changes in the magnitude of the Casimir force between two superconductors. We are able to circumvent the conventional need for complex nanopositioning, stabilization, and force detection systems. Our versatile on-chip approach is easily integrable into a dilution refrigerator allowing us to stably probe for signatures of a change in the attractive force far below the $T_\mathrm{c}$ of aluminum. In our measurements we do not observe changes in the Casimir pressure to within our resolution of 6~mPa, which, with enhanced sensitivity, will allow us to determine the validity of the Drude and plasma models. These predictions strongly depend on the specific superconducting material, resulting in varying differential forces with respect to temperature~\cite{TinkhamTc1957,Bimonte2005a,Klimchitskaya2017,EerkensPhD}. In order to properly test these predictions, the noise performance and sensitivity of our devices could be improved by increasing the length of the tethers supporting the zipper cavity or alternatively by reducing the cavity linewidth or gap size further.

The techniques developed here offer new possibilities towards fundamental physics questions that have been out of reach for conventional experiments. It is an open question how the Casimir effect changes due to the type of superconductivity (i.e., type I, II, granular, crystalline, BCS, cuprate, high temperature etc.). Our chip-based approach is compatible with a number of thin-film materials including any superconductor that can be evaporated directionally onto a substrate. One interesting prospect would be to measure Casimir forces that arise between superconductors with higher $T_\mathrm{c}$ such as lead (Pb) or cuprates such as BSCCO or YBCO.  While van der Waals stiction at sufficiently small gap sizes is a serious limitation for electromechanical devices, the unique geometry of our devices allows us to reliably revive their functionality even after structural failure due to charging (see the SI for details). The high-aspect-ratio coupling between superconductors and optics developed here could also be interesting as an optomechanics platform aiming to perform quantum microwave-to-optics frequency conversion (i.e.\ coupling zipper cavities to a superconducting LC circuit)~\cite{Bochmann2013}, which is currently a research field receiving significant attention.

\subsection*{Acknowledgements}

We would like to thank Teun Klapwijk, Holger Thierschmann, Pieter de Visser, Andrea Caviglia, Giordano Mattoni, Alessandro Bruno, and Mark Ammerlaan for their support, as well as James Quach, Markus Aspelmeyer, Ralf Riedinger, Clemens Sch\"afermeier, Dirk Bouwmeester, Wolfgang L\"offler, Kier Heeck, Galina Klimchitskaya and Alexander Krause for stimulating and helpful discussions. This project was further supported by the European Research Council (ERC StG Strong-Q, Grant No.\ 676842), the Foundation for Fundamental Research on Matter (FOM) Projectruimte grants (15PR3210, 16PR1054) and by the Netherlands Organisation for Scientific Research (NWO/OCW), as part of the Frontiers of Nanoscience program and through a Vidi grant (Project No.\ 680-47-541/994).\\


%

\clearpage

\clearpage

\setcounter{figure}{0}
\renewcommand{\thefigure}{S\arabic{figure}}
\setcounter{equation}{0}
\renewcommand{\theequation}{S\arabic{equation}}

\section*{Supplementary Information}

\subsection*{Optomechanical design and setup}
\label{SI:gOM}

\begin{figure}[ht!]
	\begin{center}
		\includegraphics[width=\columnwidth]{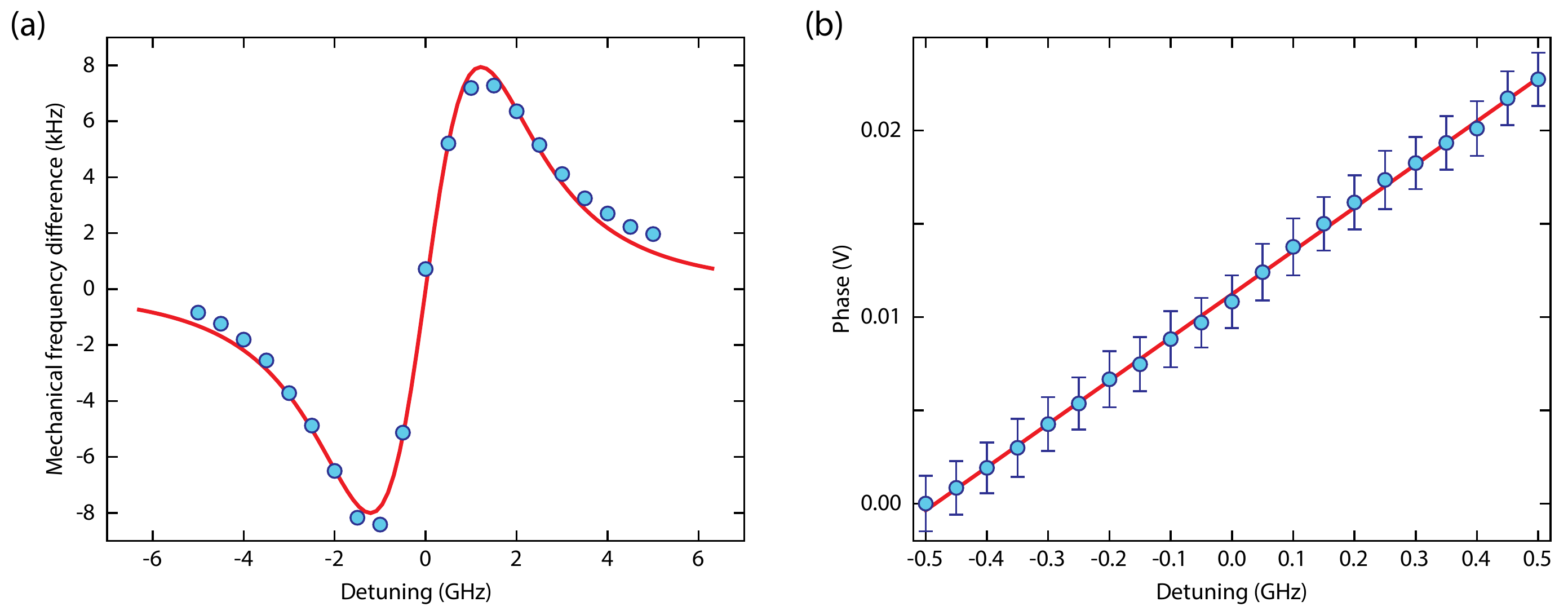}
		\caption{(a) Measurement of the optomechanical coupling rate. We sweep the detuning of the laser (input power 6.6~$\mu$W) with respect to the optical cavity and measure the response of the mechanical resonance. The fit in red is obtained form the optomechanical spring effect and used to determine $g_\mathrm{OM}/2\pi=50$~GHz/nm. (b) Calibration of the measurement sensitivity. By stepping the laser around the optical resonance and measuring the phase response we can determine our minimal sensitivity -- shown is the response between $\pm$500~MHz in steps of 50~MHz, which can clearly be distinguished. In fact our best calibrated sensitivity is around 10~MHz. The red curve is a linear fit to the data, yielding a voltage change of 0.025~mV/MHz.}
		\label{fig:S1}
	\end{center} 
\end{figure}

\begin{figure}[t!]
	\begin{center}
		\includegraphics[width=\columnwidth]{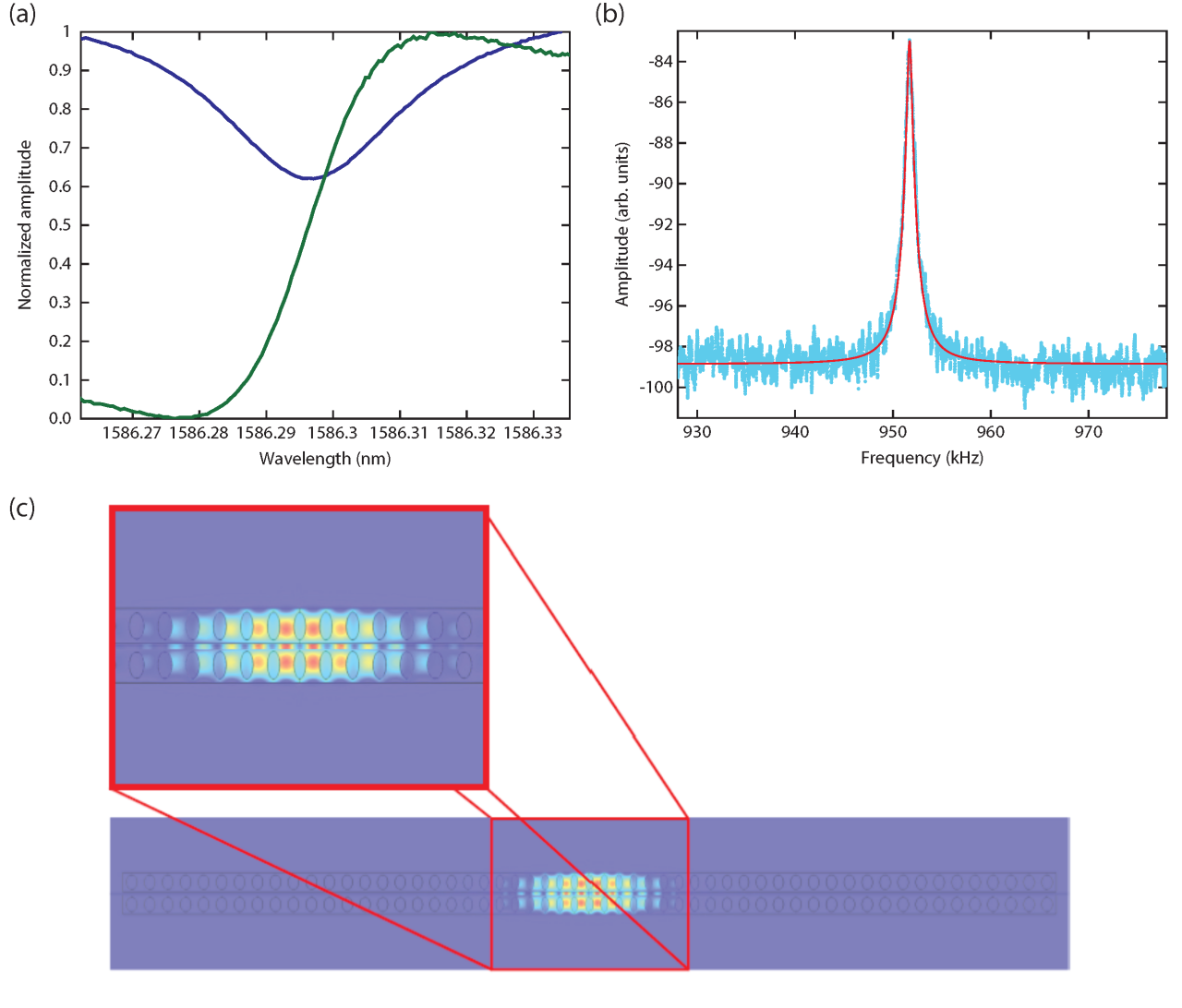}
		\caption{(a) Optical response of the cavity. Shown are the normalized amplitude (blue) and phase (green) response of the zipper cavity. The total linewidth is $\kappa = 2\pi\times 4.2$~GHz, with an extrinsic contribution of about $\kappa_\mathrm{e} = 2\pi\times 0.5$~GHz. (b) Fundamental mechanical zipper mode. The power spectral density of the mechanical zipper motion at cryogenic temperatures is plotted and fitted with a Lorentzian model (red). The measured frequency matches the simulated design very closely. (c) Finite element simulation of the fundamental localized optical mode formed between the two beams of the zipper cavity.}
		\label{fig:S2}
	\end{center} 
\end{figure}

We simulate our zipper structures using finite-element method (FEM) simulations and design it to have a fundamental cavity wavelength of 1550~nm, while the fundamental mechanical frequency of the zipper mode $\omega_m/2\pi$ is around 950~kHz. We calculate the effective mass of the mechanical mode to be $m_\mathrm{eff}=418$~pg, while the optomechanical coupling is simulated to be around $g_\mathrm{OM}/2\pi = 50$~GHz/nm. In the actual devices these parameters usually vary slightly due to fabrication imperfections and we therefore experimentally determine the most relevant ones. In particular, $g_\mathrm{OM}$ is measured by sweeping the detuning of the laser with a calibrated input power over the cavity resonance and observing the optical spring of the mechanical mode, as described in detail in reference~\cite{KrausePhD} and shown in Figure~\ref{fig:S1}(a). We also verify that all of our devices are actually freely suspended by measuring their fundamental mechanical mode, which at 4~K has a quality factor of around 2000 [see Figure~\ref{fig:S2}(b)]. The measurement of $g_\mathrm{OM}$ allows us to accurately determine the size of the gap between the two waveguides comprising the zipper cavity, which is in excellent agreement with additional measurements using a scanning electron microscope. In fact, the smallest reported gap size in the main text of 100~nm is a conservative value given these two independent measurements and we estimate the associated error to be around $\pm10$~nm.

Our setup consists of a tunable external-cavity diode laser that we couple through a lensed fiber to a SiN waveguide in the dilution refrigerator, which then couples the laser light evanescently to the photonic crystal cavity [see Figure~\ref{fig:1}]. We can adjust the optical power through a variable optical attenuator (VOA). The fiber inside the cryostat is aligned using nano-positioning stages at low temperature. The reflected light passes through an optical circulator and is detected on a photodiode. In order to achieve optimal sensitivity, we do not directly measure the optical resonance but rather the phase shift induced on the optical field by a shift in the resonance. This is realized through a Pound-Drever-Hall-like measurement~\cite{Black2001} -- we create sidebands that lie well outside the cavity linewidth by phase-modulating the laser using an electro-optical modulator at 9.8~GHz. The photocurrent from the detected light is then mixed down to DC using a mixer and a low-pass filter. Any change in resonance frequency results in a change of the level of this DC voltage [see Figure~\ref{fig:S2}]. We calibrate the sensitivity of our measurement by stepping the laser detuning around the cavity resonance using a wavemeter, which is shown in Figure~\ref{fig:S1}(b). For the mechanics measurement the photocurrent is detected on a spectrum analyzer. A sketch of the setup is shown in Figure~\ref{fig:1} and pictures of a device in Figure~\ref{fig:3}.

\subsection*{Four-point measurement}

\begin{figure}[t!]
	\begin{center}
		\includegraphics[width=\columnwidth]{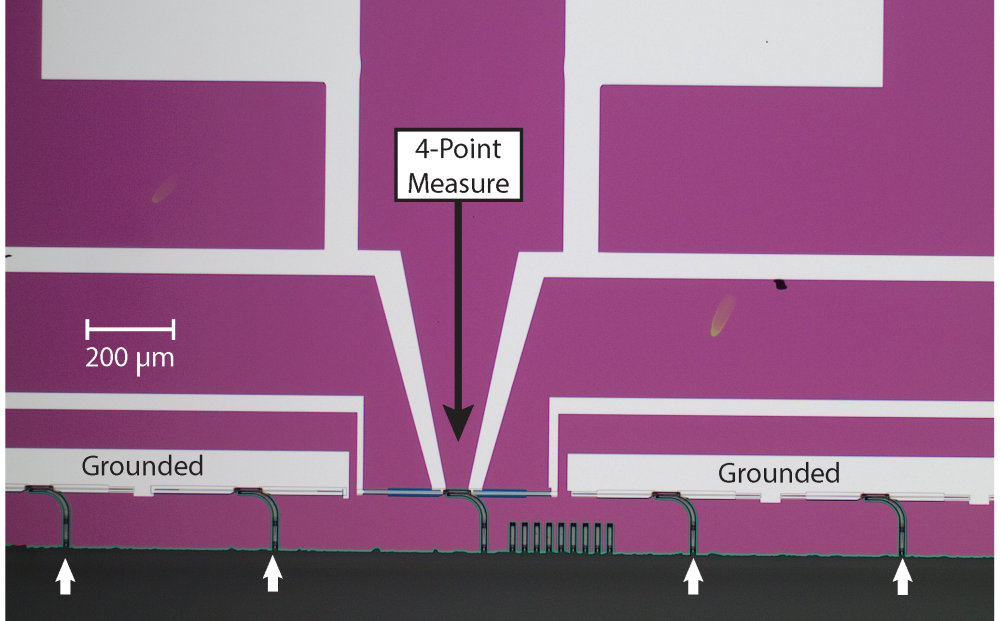}
		\caption{Microscope image showing the chip layout. The middle device is used as a reference with a four-point measurement circuit which allows us to precisely determine when the aluminum film turn superconducting. The grounded test devices (shown with white arrows) are designed to have gap sizes which we finely sweep from 100~nm to 300~nm in steps of 10~nm. This on-chip platform circumvents the need for alignment and stabilization systems conventionally required to align to different gap sizes.}
		\label{fig:S3}
	\end{center} 
\end{figure}

In order to limit the heating from the four-point measurement itself, we use a low-noise current source (Keithley 6221) to realize short, low current pulses with alternating polarity in conjunction with a nano-voltmeter (Keithley 2182A) for voltage read-out. The individual pulses have a length of 300~$\mu$s, with a delay between each pulse of 50~$\mu$s and a current of 1~$\mu$A. The noise in our resistance curves in Fig.~\ref{fig:4}(a) is an artifact of using low currents which allow us to closely estimate the heating in test devices which have no current flowing through them (and thus do not suffer from ohmic heating due to the four-point measurement). We extensively tested the length of the pulses and current level on thin, long Al wires, in order to find a regime where any residual heating from the four-point measurement can be excluded. 

When we determine the critical temperature of our superconducting film by ramping the cryostat temperature from 4~K to 10~mK, we first observe a jump in the resistivity from 45~$\Omega$ to roughly 20~$\Omega$ at a temperature of about 1.2~K. Upon further cooling, the resistivity suddenly drops to around 1~$\Omega$ just above 900~mK. These two different regimes can be explained by various parts of our chip becoming superconducting at different temperatures -- in addition to the aluminum film on the devices, which becomes superconducting at 900~mK, there is also a small part on the side of the devices coated with a superconducting film, and due to better thermal anchoring (this part is not fully undercut, unlike the devices), becomes superconducting at 1.2~K. We can clearly distinguish between these two contributions by sending significant amounts of optical power into the zipper cavity and observing a break-down of the superconductivity only along the device itself, and hence measuring a resistivity of about 20~$\Omega$.\\

\subsection*{Noise considerations}

One of the main limitations to our measurement sensitivity are drifts of the measured phase over time. Occasionally we observe long-term drifts as large as 200~MHz over the course of about 1~hr. In order to avoid these variations having a detrimental impact on our measurement sensitivity, we perform each data run within a few minutes. Such fast warm up times from 10~mK to 4~K are possible due to heaters directly attached to our mixing chamber. We carefully calibrate the fluctuations during the typical duration of a measurements and do not see spurious shifts larger than 30~MHz [cf.\ Figure~\ref{fig:S1}(b), which was measured over about 10~mins]. The origin of these phase drifts are not fully understood, and could be caused by drifts of the phase of the signal generator itself or thermal changes of the fiber lengths used in our setup, for example. Despite the sensitive nature of our measurements, we do not see any impact of other noise sources on our measurements -- for example, we can exclude seismic noise playing a role, as large vibrations caused by the pulse tube cooler of our cryostat have no observable influence on the phase measurement. Our calibrations of the read-out sensitivity [Figure~\ref{fig:S1}(b)] show that noise of the laser also does not play any significant role in our measurements.\\

\begin{figure}[t!]
	\begin{center}
		\includegraphics[width=\columnwidth]{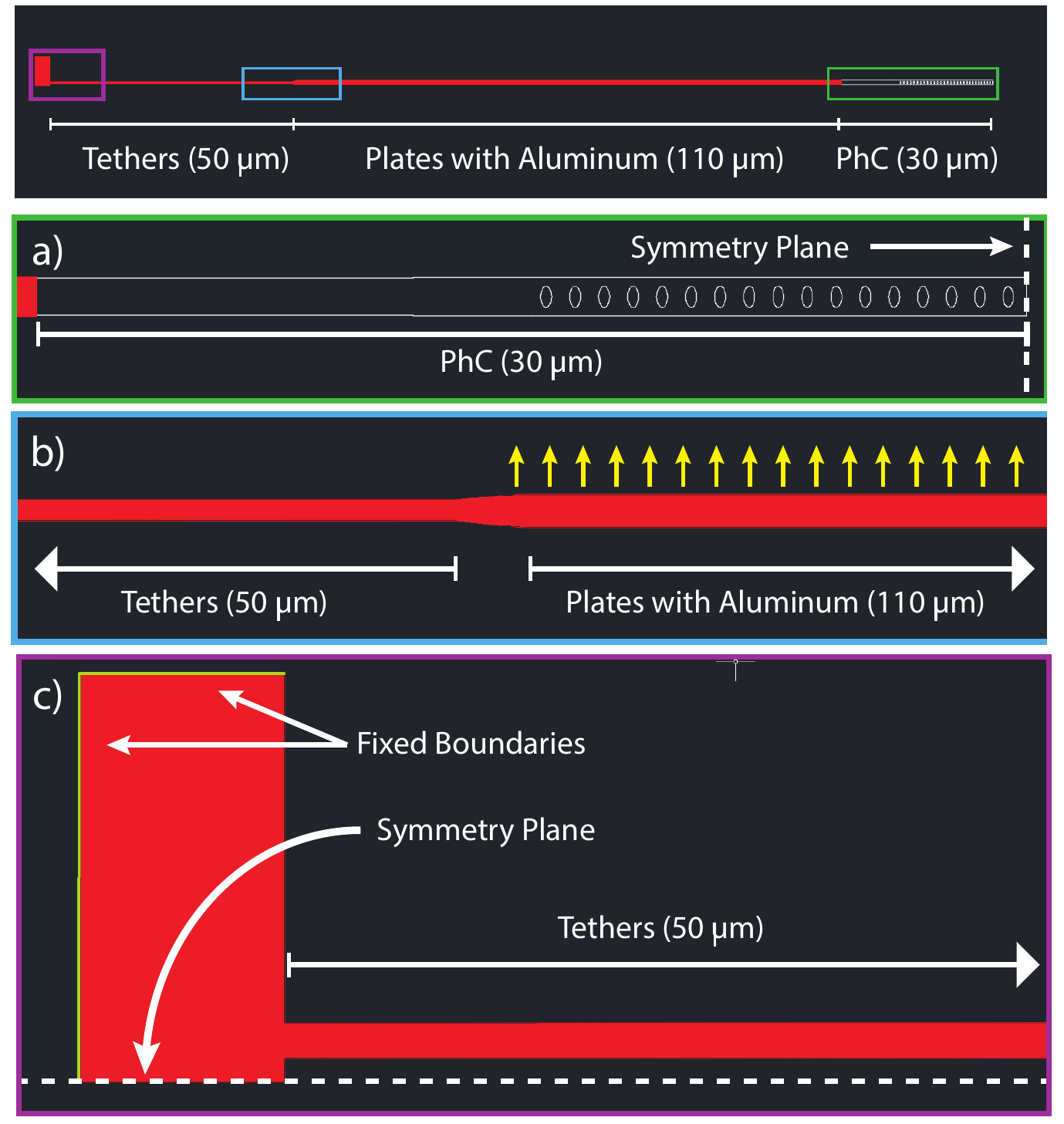}
		\caption{The top image shows the simulated structure. Since we simulate displacements of the fundamental mode, symmetry boundaries can be used to reduce the simulation size to 1/4 (i.e.\ one half of one of the two nanobeams forming the full device). The red regions indicate portions which are metalized with aluminum. In a), b) and c) we show zoomed-in views of the structure from right to left. a) A close up look of the photonic crystal cavity region and one of the symmetry planes used to cut the simulation in half. b) The metalized plate on the right forms the gap used to test for the Casimir force (the symmetric plate on the bottom is not shown). We then tether this plate into a supporting string on the left, which is 50~$\mu$m long and 50~nm wide. The yellow arrows illustrate the uniform (Casimir) pressure we simulate on the nanobeam, which gives rise to a displacement in the center of the PhC. c) Here we show a zoom-in of the clamping of the device to the substrate. The large rectangle on the left simulates the overhang that is produced during the nanofabrication process. The green borders on the top and left are the fixed boundaries of our simulations. The bottom is a symmetry boundary which simulates the other half of the zipper device.}
		\label{fig:S4}
	\end{center} 
\end{figure}

\subsection*{Aluminum film properties}

In order to calculate the expected Casimir forces between the aluminum nanowires, the coherence length and London penetration depth are required. We can determine these parameters directly using the on-chip four-point measurement. Aluminum has a bulk coherence length $\xi_0 = 1600$~nm and a London penetration depth $\lambda_L = 16$~nm. In practice, any real aluminum nanowire will be in the `dirty limit' due to unavoidable impurities in the film. Our coherence length and penetration depth must then be calculated using
\begin{equation}
\xi(T) = .85 \sqrt{\xi_0 \ell}\sqrt{\frac{T_\mathrm{c}}{T_\mathrm{c} - T}} \approx \sqrt{\xi_0 \ell}
\end{equation}
\begin{equation}
\lambda(T) = .62 \lambda_\mathrm{L} \sqrt{\frac{\xi_0}{\ell}}\sqrt{\frac{T_\mathrm{c}}{T_\mathrm{c} - T}} \approx \lambda_\mathrm{L}\sqrt{\frac{\xi_0}{\ell}},
\end{equation}
where $\ell$ is the mean free path of an electron inside the aluminum. The mean free path is determined by the geometry and quality of the aluminum nanowires. Using free electron theory, we determine $\ell$ from the measured resistance of the wires at 4~K, where the electrical conductivity is not phonon-limited. The electrical conductivity can be written as 
\begin{equation}
\sigma = \frac{1}{\rho} = \frac{n e^2 \tau}{m_{\mathrm{e}}} = \frac{n e^2 \ell}{m_{\mathrm{e}} v_{\mathrm{F}} },
\end{equation}
where $\rho$ is the resistivity, $n$ is the electron density in aluminum, $e$ is the elementary charge, $\tau$ is the mean free time between collisions, $m_{\mathrm{e}}$ the electron mass, and $v_{\mathrm{F}}$ is the Fermi velocity. We use the measured quantity $\rho\cdot\ell = m_{\mathrm{e}} v_{\mathrm{F}}/ne^2 = 4 \times 10^{-16}~\Omega$ m$^2$~\cite{Romijn1982}. One can calculate the electrical conductivity with $\sigma_{\mathrm{4K}} = L/\zeta\,R_{\mathrm{4K}}$ where $L$ is the length, $\zeta$ the cross-sectional area of the wire, and $R_{\mathrm{4K}}$ its resistance measured at 4~K. A typical value we measure is $\sigma_{\mathrm{4K}} = 4.1 \times 10^7~(\Omega\cdot m)^{-1}$, which allows us to estimate $\ell \approx 10.8$~nm. This gives an average coherence length $\xi \approx 131$~nm and $\lambda \approx 120$~nm in the `dirty limit' of our evaporated films.

The aluminum films were deposited in a thermal evaporator at $10^{-9}$~mbar with a rate of $\sim$0.3~nm/sec. The ultra-high vacuum and fast evaporation rate reduce the contamination in the films which ultimately decreases grain sizes and surface roughness. Additionally, we find that thicker depositions of aluminum (approx.\ 300~nm) dramatically increase the surface roughness of the films to about 30~nm root-mean-square. For this reason we use thin films of aluminum in our study which allow us to keep the surface roughness to $\sim$5~nm, which helps to also reduce the uncertainty in the Casimir effect~\cite{mohideen1998precision}. Unlike conventional Casimir experiments~\cite{Lamoreaux1997,mohideen1998precision} where gaps between objects are typically defined and modulated with piezoelectric stacks or electrostatic setups, our chip-based Casimir cavities have gaps which are lithographically defined and can be directly measured through optomechanical effects and imaging with a scanning electron microscope [see Figure~\ref{fig:S5}]. This allows us to accurately measure a gap variability due to parallelism and roughness of $\pm 10$~nm.

\subsection*{Finite-element simulations}

A finite-element methods simulator (COMSOL) is used to estimate the optical [see Figure~\ref{fig:S2}(c)] and mechanical properties of the metalized zipper cavities including the expected displacements at the photonic crystal (PhC) cavity due to the Casimir pressures between the metalized portions of the nanostrings. We model our fabricated silicon nitride zipper cavities with dimensions as shown in Figure~\ref{fig:S4}. The red regions indicate the parts on which we evaporate a 50~nm thick layer of aluminum. Laser interferometry measurements of the silicon substrate bending due to the tensile stress in deposited silicon nitride films show that the SiN films have 1.3~GPa of tensile stress. This tensile stress is included in our simulations and it allows us to achieve very good parallelism between the two beams which are taut even when suspended over hundreds of microns. We then simulate a uniform pressure on the side face of the metalized beams in order to estimate the expected displacement of the beams under an external pressure as shown in Figure~\ref{fig:S4}(b). Using the stress map of the beams as an initial condition for simulating the mechanical modes of the structure, we calculate a fundamental mode frequency of 950~kHz, in good agreement with the measured frequency of 952~kHz [cf.\ Figure~\ref{fig:S2}(b)]. This mechanical mode is the fundamental differential mode of the two beams.

\subsection*{High-aspect-ratio parallelism}
\label{SI:charge}

\begin{figure}[t!]
	\begin{center}
		\includegraphics[width=\columnwidth]{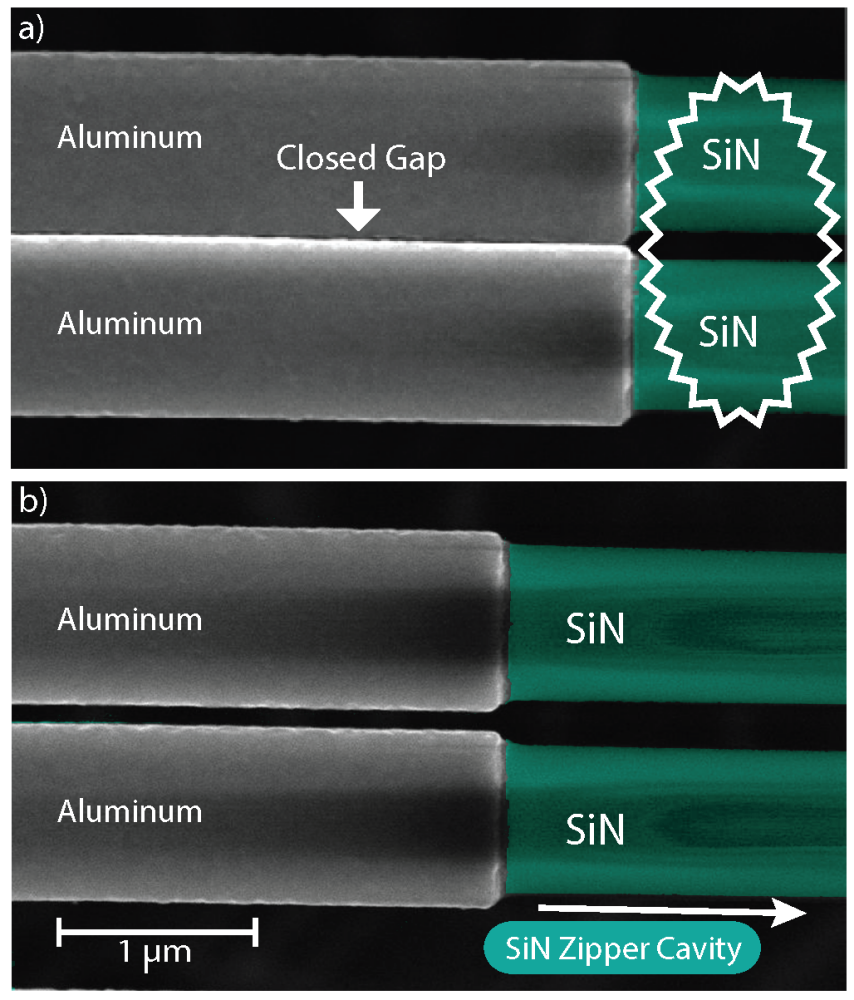}
		\caption{a) Scanning electron microscope (SEM) image showing a close-up look at a device which has stuck together due to static charges induced during post-fabrication handling. An SEM can be used to charge and separate the un-metalized portions of the SiN nanobeam. b) Shown is the same device as in a) completely opened after purposefully charging with the SEM. After the nanobeam dissipates charge through its clamping, the stress in the SiN dominates, pulling the gaps back together to the target spacing.}
		\label{fig:S5}
	\end{center} 
\end{figure}

One of the primary challenges in Casimir experiments is concurrently achieving excellent parallelism between two large surfaces with the smallest possible spacing between them. This is also a major obstacle in reliably fabricating nano-structures with closely-spaced, suspended components needed for on-chip Casimir experiments and optomechanical devices. With increased aspect-ratios and flexibility, these metalized components can become highly susceptible to stiction caused by static charging during handling [Figure~\ref{fig:S5}(a)]. Once stuck together, van der Waals forces usually keep these gaps permanently closed, rendering the devices unusable for experiments. This points to a trade-off between high-aspect-ratio parallelism and fabrication yield. However our platform's unique geometry allows us to circumvent this well-known bottleneck. Once fabricated, we can image our devices for signs of stiction using a scanning electron microscope (SEM). The long, suspended geometries of our SiN nanobeams also mean they can be easily charged using an SEM by magnifying on the un-metalized portions of our SiN structures as seen on the right of Figure~\ref{fig:S5}(a). With enough charging, we can induce a large separation between the SiN beams (on the order of microns) which also pull open the aluminum-covered portions of our devices. As the nanobeams slowly discharge via clamping to the substrate, the high stress in the films begins to dominate and pulls the beams together to their targeted gap size as seen in Figure~\ref{fig:S5}(b). This ability to revive the functionality of our devices allows us to push the limits of achievable high-aspect-ratio nano-gaps between superconductors.

\def\urlprefix{}

\end{document}